# Augmented Reality for Education: A Review

Carlo H. Godoy Jr.
College of Industrial Technology,
Technological University of the Philippines
Manila, Philippines

**Abstract:-** Augmented Reality, or simply AR, is the incorporation of information in digital format that includes live footage of a certain user's real-time environment. Also now, various universities are using Augmented Reality. Applying the technology in the education sector can result in having a smart campus. In line with that, this paper will discuss how Augmented Reality is being used now in different learning areas.

## I. INTRODUCTION

According to Rouse as cited in [1], Augmented Reality or simply AR is the integration of information in digital format which includes live video on the real time environment of a certain user. In an augmentation of live videos, integrating a video picture to digital environment involves identification of an object replicated from the physical world features and will be captured as any format that will be considered as a video picture which will mean that increasing the responsiveness of the generated video picture to the state needed to control the object from the physical world itself [2] In an augmented reality system, the integrated digital information can only be seen using a medium like phone cameras but it will not be seen in the real world. These digital information can be represented in different forms like a stack of virtual cubes or manipulating a non-real object in many ways possible [3].

Another kind of integration that Augmented Reality is normally defined for is it can indicate a supplemental information to a user. This supplemental information are considered optional and may not affect the actual user of the system itself. The method being used by an Augmented Reality System to provide these supplemental information are the following: Tracking the user's point of view, Capturing a camera field of perspective and Obtaining additional data in the field of perspective captured for at least one object [4]. One perfect example of this supplemental information which obtains additional data in the field of perspective captured for at least one object is if the user's interest is in vehicle, the system should present an augmented reality replica of a vehicle and cover the user's environment based on the point of interest [5].

## II. AUGMENTED REALITY FOR EDUCATION

Augmented Reality (AR) apps have received increasing attention over the previous two decades. In the 1990s, AR was first used for apps linked to pilot education as well as for training of Air Force [6]. AR generates fresh world experiences with its data layering over 3D space, suggesting that AR should be embraced over the next 2–3 years to give fresh possibilities for teaching, learning, study, or creative investigation according to the 2011 Horizon Report [7]. AR uses virtual objects or data that overlap physical objects or environments to create a mixed reality in which virtual objects and actual environments coexist in a meaningful manner to increase learning experiences. Azuma as cited by [6], stated that the mentioned virtual objects is appearing in coexistence as the same space as the objects that is located in the real world. AR is now a common technology commonly used in instructional environments in the education sector [8].

AR has also become a major study focus in latest years. One of the most significant factors for the widespread use of AR technology is that it no longer needs costly hardware and advanced machinery such as head mounted screens [6]. According to Chiang, Yang & Hwang as cited by [6], Augmented Reality is widely used now in the K-12 level in the education industry. Ferrer-Torregrosa et al., (2015) stated that Augmented Reality is also being used now by different universities. The application of the technology in the education sector can lead to have a smart campus. Smart campuses are designed to benefit professors and students, handle the resources available and improve the experience of the users with proactive services [10].

A smart campus is normally designed for smart cities. A smart city is a city infrastructure that includes technological design as remedy for problems faced by its citizens. In a smart city, every structure starting from information systems to transport technologies, from libraries, hospitals and schools, to other community services are modeled technologically [10]. These community services include linkages of the students with environmental context awareness. Augmented reality (AR) provides potential benefits for increasing understanding of environmental context awareness and the cultural framework and increasing the experiences of learners in live environment settings through strategically combining electronic things with a live environment setting [11].

As highlighted in the Horizon report, Augmented Reality (AR) is acknowledged as one of the most significant innovations in greater and K-12 education technology [12]. Augmented reality is gradually becoming integrated as an emerging technology in the region of inclusive education that adapts learning in equal footing through exploration and experience by all [13]. Johnson (as cited by Saltan, 2017) stated that AR is anticipated to be widely adopted in higher education for two to three years and in K-12 for four to five years. It is essential to explore how teachers and scientists incorporate AR into teaching-learning procedures if this is the present state of the art for the use of AR in education. MUVEs and AR became visible in the early 2000s and their effectiveness for learning was soon established by educational research [15].

## III. RESEARCH METHODOLOGY

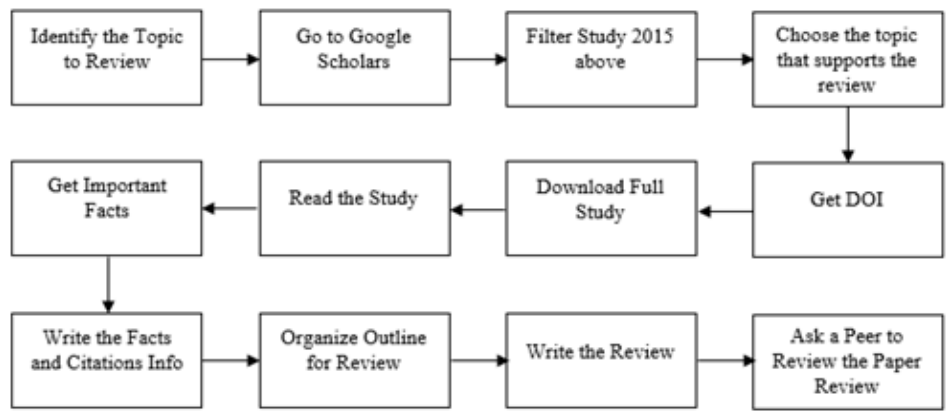

Fig 1:- Overview of Research Methodology

The searching procedure started by selecting the topic to be reviewed. In this case the topic is Augmented Reality for Education. The topic selected will explore the different sectors in Education that is using Augmented Reality as a tool for teaching and learning. After identifying the topic, the next step is to go to google scholar. In this part the study will be filtered depending on the importance of each study. Another filter that needs to be added is the year when the publication was published. It is very safe to say that five years interval will still make a certain publication still valid. Once the filtering has been set, it is now time to select the relevant document that will build up the foundation of the review. After knowing the foundation and the outline where the documents will be discussed, the DOI will be used to be able to get the full copy of the documents. Once the full study has been acquired, it is now time to review all documents. During the review process, this is the time to filter which documents is needed to support the selected topic. Take down notes and get all the helpful information for the citation. Once the important facts has been gathered and the studies has been filtered, it is now time to write the review.

## IV. AUGMENTED REALITY FOR EDUCATION: A REVIEW

*A. Augmented Reality for English Education.*

Recently, the world has focused on reading understanding as a consequence of the International Student Assessment Program (PISA), which stated that only about 8% of learners in OECD nations are top readers [16]. Including digital technology like an Augmented Reality Technology inside the classroom is a precious instrument for achieving required norms and promoting convincing result by the experiences of the students involving learners in learning activities including reading. Comprehensible written and oral input is important for students enrolled in langueage education due to the fact that usage of animations, sounds, videos and pictures enriches the entry and long-term and exciting learning of those students [17].

In this regard, AR technology provides many language teaching and learning possibilities. An Augmented Reality based game called ChronoOps has been used to scientifically test the behaviors of language learners. An scientific study of language students involved in using an AR location-based portable match that introduces situational as well as encouraging respondents to grow beyond the traditional subject roles connected with ' student ' or ' learner ' roles is the focus of the ChronoOps [18].

*B. Augmented Reality for Foreign Language Education.*

Initially, Arabic language teaching can no longer depend exclusively on traditional learning techniques such as note-taking and lecturing, which are still predominantly preferred among Arabic lecturers as Zainuddin & Sahrir (2016) stated. Early evaluation and observation by the scientists means that the absence of usage in instructional digital advancements for learning as well as teaching Arabic has hindered the memorization process of teaching Arabic vocabulary in the classroom. Ismail (as cited by Zainuddin, Sahrir, Idrus, & Jaafar, 2016) addressed and suggested that an emerging needs like this requires an intervention of using an educational systems in Arabic linguistic teaching such as Arabic courseware.

Educators and skilled trainers should handle its execution, with comprehensive knowledge and comprehension of the inherent and extrinsic motivations of the learner, thus creating a good personalized atmosphere. Increased truth has unexplored educational potential and capacity to help learners seamlessly in a natural environment. For Radu (as cited by Zainuddin, Sahrir, Idrus, & Jaafar, 2016) Augmented Reality works as an e-learning instrument for enhanced comprehension of content, learning spatial constructions, language connections, long-term memory retention, enhanced cooperation and motivation. In terms of french language, a learning tool called Explorez has been developed. Explorez enables learning to happen outside the classroom with the objective of offering a contextual and immersive educational experience: one that is important and applicable to the students [21].

*C. Augmented Reality for ICT Education.*

In instructional environments, computer techniques were implemented and made learning more flexible and intuitive. Augmented reality (AR) has attracted great government attention among these techniques because it offers a fresh teaching view by enabling learners to visualize complicated spatial relationships and abstract ideas [22]. Research have shown that, owing to a number of factors, many Malaysian non-technical learners have low motivation in studying ICT courses, such as absence of teaching practice and efficient teaching apps. In the outlook of such issue, the research teams conducted a quasi-experimental analysis to examine the adverse effect of a new application for mobile augmented reality learning (MARLA) on the motivation of learners to learn a topic of an ICT course at the university [23].

The analysis showed more motivation for masculine learners than for their inverse counterparts. In addition, there was an impact of interaction between gender and method of learning, with male learners achieving distinct motivation levels depending on technique of learning. Such a mobile learning instrument may possibly be used to assist non-technical undergraduates learn more in a motivated manner, but their achievement will depend on adequate planning and execution by considering the demographic background of the learners.

Another study under the ICT education sector has a goal in exploring if the integration of AR methods would facilitate application for changing the style as well as analyzing a distinct outcome in educating the learners which uses blended learning approach based on online and AR [24]. It was found that technology instructional scientists should take cautious consideration of the educational goal architecture, the data size shown on the cellphone monitor and the teaching machinery and school facilities setting when incorporating AR apps into a course in order to obtain an appropriate learning situation.

*D. Augmented Reality for Science Education.*

Education Professionals must tackle several problems intrinsic in the training of science fields such as physics– costly or inadequate laboratory equipment, mistake of equipment, difficulty in simulating certain experimental circumstances [25]. Augmented Reality (AR) can be a successful approach to tackling these problems. A study about magnetic field instruction has been conducted in relation to the aforementioned problems. Results of the analysis demonstratesd that the movement-sensing software based on AR can enhance the learning attitude and learning result of the learners. This research offers a case for applying AR technology to secondary education in physics [25].

In learning about health science, medical anatomy and neurosurgical it is also very helpful to use Augmented Reality as a learning tool. In an environment where required structure needs to be examined from all angles, anatomical learning is best performed using a tool that will show this angles [26]. Augmented Reality is one of the best tool to show angles as the developer can easily manipulate how the augmented object will rotate and show. Compared to traditional pedagogical schemes, VR and AR have the ability to produce improved teaching environments. 3D learning environments can increase the motivation / engagement of learners, improve the representation of spatial information, improve learning contextualization and create superior technical skills. Over the previous several centuries, neurosurgical has experienced a technological revolution, from trephination to image-guided navigation. Advances in Virtual Reality (VR) and Augmented Reality (AR) are some of the latest ways of integrating into neurosurgical exercise and resident education [27].

Studies have shown that AR technology can significantly improve the results of education. For example, AR enables learners participate in real-world genuine explorations such as marine life explorations that not everyone has been able to achieve [28]. Marine schooling includes problems that are wealthy and multifaceted. Raising awareness of marine settings and problems requires fresh teaching materials to be developed.

In line with that, a digital game-based learning was tailored for primary school learners to design an innovative marine learning program incorporating augmented reality (AR) technology [29]. The results of using this technology are the following: (1) learners were extremely confident and satisfactorily viewed the learning operations ; (2) learners obtained target goal for understanding ; and (3) the innovative teaching program specifically helps small academic achievements and enhance learning efficiency. Another great application of Augmented Reality in science is an AR-based simulation scheme for a cooperative investigation-based teaching activity in a science course and discovered that AR-based simulation could involve learners more deeply in the investigatory project activity than traditional simulation could [30].

*E. Augmented Reality for Social Science and History Education.*

According to Field Day Lab (2016), an expedition leader can be compared to one of the many roles a teacher plays. Teachers are leading their learners on a discovery trip that extends their knowledge of the globe around them and prepares them to become more knowledgeable, curious, perhaps even more empathic citizens of the globe. Simulation, immersion, and cultural learning attract researchers from a multitude of areas including anthropology, cognitive psychology, company, and education.

Cultural learning in particular is closely linked to learning languages because language is the main component of cultural contexts and students cannot really master the desired language until they have understood cultural contexts as well. In cultural and language teaching, physical-virtual immersion and real-time communication play an important role. Augmented reality (AR) technology can be used to fuse virtual items seamlessly with real-world pictures for immersion [32]. Adding augmented reality to distant interaction implies that individuals can communicate with other individuals or items without being there physically. We leave the prehistoric cavern paintings, the painting work of the panoramists, the photographers and videographers behind to lastly' join the picture [33]. If we follow the evolution of the representation device, we can see that we are entering the era of' frameless pictures,' pushing us back to square one that is how Augmented Reality is changing the norms of teaching and learning social science and history.

*F. Augmented Reality for SPED Learning.*

Disabled learners are growing substantially and demographically worldwide, yet distance schooling has failed to tackle their particular requirements to generate a completely inclusive educational experience. Augmented Mobile Reality and its separate features provide a chance to remedy this condition [34]. A research investigated a fresh way for kids with distinct disabilities to incorporate sophisticated display technology into instructional operations. A free interactive portable augmented reality (AR) application was created to promote the teaching of geometry. Twenty-one kids from elementary school took part on the research. As the findings suggest, Augmented Reality scheme would aid college kids complete puzzle game tasks regardless of teacher support.

Using AR display technology, respondents showed enhanced capacity in completing puzzle related assignments contrasted to document-based traditional techniques. As information on achievement through labor stated, usage of Augmented Reality application in kids with special needs could increase learning motivation and tolerance for frustration [35]. Another Augmented Reality based application has been create for special needs education called "Fancy Fruits." It is used to teach children with disability the components of regional vegetables as well as regional fruits. The app contains marker-based AR components that connect with virtual data to the actual scenario. A field survey was carried out to assess the request. The research was attended by eleven kids with mental disabilities. The findings indicate that the respondents has a high level of pleasure [36]].

*G. Augmented Reality for Vocational Training Education.*

Azuma (as cited by Yilmaz, 2016) stated that Augmented Reality is described as having the following characteristics: integrating actual live environment with computer created environment, offering conversation as well as showing 3D items. All of the mentioned components can really be helpful to develop psychomotor skills of vocational trainees through simulation method. By using simulators, trainees can easily replicate the methodologies of a certain industrial based training. In TVET organizations, educators sees significant challenges on learning system owing to a broad range of SPED necessity of learners a. A marker-based mobile Augmeted Reality app called Paint-cAR has been created in aiding the method of teaching fixing car paint as included in vehicle maintenance vocational training program [38]. The application was created using a methodology and principle of UDL to aid or assist deeply in the development of portable augmented apps in instructional Collaborative creation purposes. To validate Paint-cAR application in a true situation, a cross-sectional assessment survey was performed.

*H. Augmented Reality for Mathematics Education.*

An integrated STEM (Science, Technology, Engineering and Mathematics) lesson requires to participate and nurture students ' interest in real-world circumstances. While real-world STEM situations are naturally incorporated, the embedded STEM contents are rarely taught by school educators [39]. One of the hardest subject of that track is Mathematics. One example of a Mathematics subject is Solid Geometry. To give a better experience in learning solid geometry, a study has been conducted to combine Augmented Reality (AR) technology into teaching operations designing a learning scheme that helps junior high school learners learn sound geometry [40], [41]. Based on the result of the study, AR really gives a big leap in learning solid geometry.

Another study deals with the use of AR in teaching and learning math that uses this technology to its complete benefit in providing concrete experience in interacting with revolutionary solids. At the end of the study, it was found out that Augmented Reality is beneficial in the understanding of computing solids of revolution volumes [42]. AR techniques are strongly linked to calculation capacity and computational calculations, and therefore their evolution is related to personal computer development. It is therefore essential to begin by referring to some of the works that have been created through the implementation of these techniques at global and national level, primarily in the field of education and teaching [43].

With, it can easily be inferred that from the birth of AR it is already related to Mathematics. AR makes mathematical ideas simpler to comprehend because it provides better visualization and interaction. We can therefore conclude that three-dimensional techniques, such as AR, improve mathematics teaching and learning. At the same time as the imperative to better comprehend the use of mobile devices for learning mathematics in many nations, there is a powerful political will to enhance teaching and learning process in mathematics education to support innovation that drives economic growth and create the capacity of tomorrow's workers for future work markets [44].

## V. CONCLUSION

Research has shown that AR can be more efficient in supporting teaching than other improved settings in technology. If content is represented as 3D learners, objects can be manipulated and information handled interactively (El Sayed, Zayed, & Sharawy as cited by Buchner & Zumbach, 2018). Rapid technological evolution has altered the face of education, particularly when technology has been coupled with appropriate pedagogical foundations. This combination has developed fresh possibilities to enhance teaching and learning experience quality [46].

Based on the findings, Augmented Reality (AR) is a technological strategy that offers apps that enable learners to communicate with the actual globe through virtual data, and Game-Based Learning (GBL) is a pedagogical strategy that promotes the use of learning games to sum up all preceding discussions. Combining the two process will definitely result to a new system that will give a big impact in the education industry [22].